\documentclass[12pt,preprint]{aastex}

\shorttitle{AN EXTENDED STAR CLUSTER AT THE OUTER EDGE OF M\,33}
\shortauthors{STONKUT\.{E} ET AL.}

\received{2007 August 23} \accepted{}

\begin{document}

\title{AN EXTENDED STAR CLUSTER AT THE OUTER EDGE \\
       OF THE SPIRAL GALAXY M\,33 \altaffilmark{1}}

\author{Rima Stonkut\.{e}\altaffilmark{2}, Vladas
Vansevi\v{c}ius\altaffilmark{2}, Nobuo Arimoto\altaffilmark{3,4}, \\
Takashi Hasegawa\altaffilmark{5}, Donatas Narbutis\altaffilmark{2},
Naoyuki Tamura\altaffilmark{6}, \\ Pascale Jablonka\altaffilmark{7},
Kouji Ohta\altaffilmark{8}, and Yoshihiko Yamada\altaffilmark{3}}

\altaffiltext{1}{Based on data collected at Subaru Telescope,
which is operated by the National Astronomical Observatory of
Japan} \altaffiltext{2}{Institute of Physics, Savanori\c{u} 231,
Vilnius LT-02300, Lithuania} \altaffiltext{3}{National Astronomical Observatory of Japan, Mitaka, Tokyo 181-8588, Japan} \altaffiltext{4}{Department of Astronomy, Graduate University of Advanced Studies, Mitaka, Tokyo 181-8588, Japan} \altaffiltext{5}{Gunma Astronomical Observatory, Agatuma, Gunma 377-0702, Japan} \altaffiltext{6}{Subaru Telescope, National Astronomical Observatory of Japan, 650 North A'ohoku Place, Hilo, HI 96720, USA} \altaffiltext{7}{Universit\'e de Gen\`eve, Laboratoire d'Astrophysique de l'Ecole Polytechnique F\'ed\'erale de Lausanne (EPFL), Observatoire, CH-1290 Sauverny, Switzerland} \altaffiltext{8}{Department of Astronomy, Kyoto University, Kyoto 606-8502, Japan}

\begin{abstract}
We report a discovery of an extended globular-like star cluster,
M\,33-EC1, at the outer edge of the spiral galaxy M\,33. The
distance to the cluster is 890\,kpc, and it lies at a 12.5\,kpc
projected distance from the center of M\,33. Old age ($\gtrsim$7\,Gyr)
and low metallicity ([M/H]\,$\lesssim$\,-1.4) are estimated on the basis of isochrone fits. Color-magnitude diagrams of stars, located in the cluster's area, photometric and structural parameters of the
cluster are presented. Cluster's luminosity ($M_{V}$\,=\,-6.6) and
half-light radius ($r_{\rm h} = 20.3$\,pc) are comparable to those
of the extended globular clusters, discovered in more luminous
Local Group galaxies, the Milky Way and M\,31. Extended globular
clusters are suspected to be remnants of accreted dwarf galaxies,
and the finding of such a cluster in the late-type dwarf spiral
galaxy M\,33 would imply a complex merging history in the past.

\end{abstract}


\section{Introduction}

Resolved stellar diagnostics has been extensively applied for
investigation of merging history of galaxies. In this context
extended stellar systems have been recently known to be
informative. Firstly, some of the extended stellar systems in the
Milky Way (MW), e.g., M\,54 and $\omega$Cen, are suggested to be
remnants of accreted dwarf galaxies, which might be responsible
for the thick disk and halo formation. Such systems have produced
large-scale stellar streams in the MW, thus they are useful to
highlight various substructures of the host galaxies and to reveal
their merging history. Secondly, while the key physical processes
that discriminate extended star clusters and low surface
brightness dwarf spheroidals (dSphs) are poorly understood, their
distinction is rather ambiguous.

Searches for extended stellar systems discovered at least a dozen
of low surface brightness dSphs in the vicinity of MW \citep{sak06, bel07, irw07} and M31 \citep{mar06}. Recently \citet{hux05} and
\citet{mac06} discovered four extended luminous star clusters in
the vicinity of M\,31.  Star clusters of this type are also found
in the spirals M\,51 and M\,81 \citep{cha04}, and in the giant
elliptical galaxy NGC\,5128 \citep{gom06}. It is important to
stress, however, that all extended star clusters found so far
belong to massive luminous galaxies. Therefore, any piece of
evidence on extended stellar systems in smaller galaxies would
play an important role in disclosing the merging history of
galaxies on all scales.

M\,33 is a unique late-type (Scd) dwarf spiral galaxy in the Local
Group, resolvable by ground-based observations, and it is claimed
\citep{fer06} to possess an unperturbed stellar disk without any
remarkable sign of thick disk and halo. However, \citet{cha02}
revealed an old cluster population, which has a velocity
distribution that they attributed to the thick-disk/halo
component. Warp of the M\,33 gaseous disk has already been known
from HI observation \citep{cor89}, and a stellar stream was
suggested recently from spectroscopy of individual stars \citep{mcc06}. Any further evidence on the M\,33 perturbation and accretion
events is indispensable to disclose the real formation history of
the galaxy. Since stellar systems of accretion origin are often
found far from the hosts' central part, wide and deep searches for
such objects are crucial.

We report a discovery of an extended star cluster, M\,33-EC1, in
the M\,33 photometric survey (P.I. N.\,Arimoto) frames obtained on
Subaru Telescope (Fig.\,1). The cluster is located at
R.A.\,=\,$01^h32^m58\fs5$, Decl.\,=\,29\degr 52\arcmin 03\arcsec\
(J2000.0), lying far south from the M\,33 center at a projected
galactocentric distance of 48\farcm4. Previous M\,33 cluster
studies did not reveal any clusters of a comparably large size
(\citealp{cha99}, 2001). An extensive catalogue of M\,33 star
clusters recently compiled by \citet{sar07} does not include this
new object.

In section 2 we present details of observations and data
reduction. In section 3 the derived cluster parameters and
resolved stellar photometry results are given. In section 4 we
briefly discuss the impact of our finding in the context of galaxy
formation.

\section{Observations and Data Reductions}

Photometric data of the discovered star cluster, M\,33-EC1, were
obtained during the course of the M\,33 wide field photometric
survey performed on Subaru Telescope, equipped with Prime Focus
Camera (Suprime-Cam; \citealp{miy02}). Single shot Suprime-Cam
mosaic (5$\times$2 CCD chips; pixel size of 0\farcs2) covers a
field of 34\arcmin$\times$27\arcmin, and a magnitude of
$V\sim25^m$ is reached in 60\,s. Broad-band images: $V$-band
(exposures 5$\times$90\,s; seeing $\sim$1\farcs0), $R$-band
(5$\times$90\,s; $\sim$0\farcs6), and $I$-band (5$\times$200\,s;
$\sim$0\farcs8) were acquired during photometric nights. For
standard reduction procedures we used the software package
\citep{yag02} dedicated to the Suprime-Cam data. We employed the
DAOPHOT \citep{ste87} program set implemented in the IRAF software
package \citep{tod93} for crowded-field stellar PSF (point spread
function) photometry and integrated aperture photometry of the
cluster. The PSF stellar photometry on 5 individual exposures in
each passband was performed.

Instrumental magnitudes were transformed to the standard
photometric system by referring to the published M\,33 photometric
catalogue \citep{mas06}. In total 220 stars spanning the $I$-band
magnitude range from 19$^m$ to 21$^m$ and wide color ranges ($R-I$
from -0.15 to 1.3; $V-I$ from -0.25 to 2.5) were selected as local
standards. R.M.S. errors of the transformation equations for $V-I$ and $R-I$ colors, and the $I$-band are less than 0$\fm035$ which, taking into account the number of employed stars, assures accurate calibration. Considering the intrinsic calibration accuracy of the standard stars \citep{mas06}, we estimate the accuracy of
our photometric data to be of $\sim$0$\fm015$ at $I=22^m$.
We used a bilinear $R-I$ color transformation equation due to a
significant difference between the transmission curve of the Suprime-Cam $R$-band interference filter and that of the standard Cousins $R$-band filter.

The star cluster M\,33-EC1 is located far beyond the M\,33
galaxy's disk, therefore, it is reasonable to assume that its
colors are contaminated only by the MW's foreground extinction.
Photometric data were de-reddened using the $E(B-V)=0.06$ value,
derived at the cluster's position from the extinction maps
\citep{sch98}, as follows $A_V=3.1\cdot E(B-V)$, $A_I=0.11$,
$E(R-I)=0.045$.

\section{Results}

\subsection{Color-magnitude diagram}
The color-magnitude diagram (CMD) of a region of 20\arcsec\
radius, centered on the star cluster M\,33-EC1, is dominated by
red giant branch (RGB) stars, see Fig.\,2. Reduction
and photometry procedures enable us to recognize and remove
obvious bright non-stellar objects (star/galaxy separation was
performed by eye referring to PSF fitting parameters --
sharpness and $\chi^2$), however, faint unresolved background
galaxies can still be present in this diagram.

In order to resolve well-known age-metallicity degeneracy of the
RGB position in CMD, inherent to old populations, it is helpful to
introduce faint RGB and horizontal branch stars into the isochrone
fitting procedure, see e.g., \citet{mar06}. The global shape of
our CMD resembles the CMD plotted in Fig.\,7 from \citet{mar06},
implying the presence of a very old population with a prominent
horizontal branch. However, the limiting magnitude of our
observations is too shallow for reliable morphology study of the
lower part of the CMD. Therefore, to estimate the intrinsic RGB
width over the entire magnitude range, and to derive radial and
magnitude dependence of data completeness, we performed an
artificial star test (AST) on $R$- \& $I$-band images. The AST
results quantify in detail the photometry errors, confusion limits
and data completeness, making the isochrone fitting procedure more
robust and better constrained.

Six reference points on the observed RGB ($I,R-I$\,=\,20.90, 0.72;
21.90, 0.65; 22.90, 0.57; 23.40, 0.53; 23.90, 0.49; 24.40, 0.46)
were selected to represent the entire magnitude range of the
cluster's stellar population. DAOPHOT's {\it addstar} procedure
was employed to add artificial stars to the images. To avoid
self-crowding we generated individual AST images at every
reference point. Each AST image contains 400 artificial stars of
the same magnitude distributed on a regular grid (step 3\arcsec)
over the region of 60\arcsec$\times$60\arcsec\ centered on the
cluster. However, only 140 artificial stars fall within the actual
cluster radius of 20\arcsec. In order to increase the number of
artificial stars and derive radial data completeness distributions
more reliably, we generated 21 individual images for each passband
and every reference point by shifting the grid around the initial 
position to 8 and 12 symmetrically distributed locations around 
the initial position at the radial distances of $\sim$0\farcs6 and 
$\sim$1\farcs2, respectively. Therefore, within the
radius of 20\arcsec\ we used 2940 artificial stars in total at
each reference point on the RGB. The photometry procedure of the
AST images was exactly the same as the one employed for the real
star photometry.

To understand the morphology of star distribution in the lower
part of CMD we constructed artificial star CMD. Radial
distribution of the artificial stars at every reference AST point
on the RGB was chosen to represent the observed radial density
distribution of the cluster stars. However, to increase robustness
of the artificial star CMD, we used a number of artificial stars 5
times greater than the number of real stars. The observed cluster
stars over-plotted on the artificial star CMD are shown in Fig.\,2, panel b).

``Christmas tree-like'' artificial star CMD (Fig.\,2, panel b)
implies that CMD of the star cluster M\,33-EC1 is composed solely of
RGB stars, experiencing very low contamination by foreground stars
and background galaxies. Note, however, the enhanced (in respect to
the artificial stars) density of the faint blue ($R-I<0.25$)
objects, which could be attributed to horizontal branch stars of
the cluster or faint blue galaxies. Therefore, the straightforward
isochrone fit to the observed stars can be applied down to
$I=23^m$, using only the RGB part of the isochrones.

We constructed the radial data completeness plot (Fig.\,3) by
counting the recovered artificial stars in 2\arcsec\ wide annulus
zones centered on the cluster. Stars down to $I=23^m$ are well
recovered even at the very center of the cluster. At this
magnitude level we are able to find and measure more than 70\% of the
stars at the cluster's center and more than 95\% at larger radii (Fig.\,3).

The $\sim$100\% data completeness of the brightest RGB stars is of
high importance for the cluster's distance determination by
fitting the tip of RGB (TRGB). This method is based on the
assumption (valid for [Fe/H]\,$\le$\,-0.7 and ages of
$\gtrsim$2\,Gyr) that the absolute $I$-band magnitude of TRGB
($M_I=4.05\pm0.10$) is independent of metallicity and age
(\citealp{lee93}; \citealp{bel01}). The AST data completeness
results imply that the TRGB method can be applied throughout
the radial extent of the star cluster.

A magnitude of the brightest RGB star (it is located within the
cluster's core, however, in an uncrowded area, and thus measured
accurately) is of $I=20.81\pm0.01$. Taking into account the
MW foreground extinction ($A_I=0.11$), this converts to a distance
modulus of $(m-M)_0=24.75\pm^{0.10}_{0.20}$ and places the star cluster M\,33-EC1 at a distance of $890\pm^{40}_{80}$\,kpc. The distance modulus error is dominated by the systematic error of the TRGB calibration ($\pm0.10$) and by an additional increase of the distance modulus, arising due to a probability, that the brightest observed star is below the very tip of theoretical RGB, because of a small total number of RGB stars in the cluster.

It is noteworthy to stress, that we determine the RGB tip of the
M\,33 galaxy's outer disk at $I=20.68\pm0.02$, which converts, by
applying the MW foreground extinction, $A_I=0.08$, and assuming validity of the TRGB method for the case of M\,33 outer disk's metallicity, to $\sim$850\,kpc. The derived distance of M\,33 is in agreement with recent M\,33 galaxy distance determinations, based on the TRGB method, by \citet{gal04} and \citet{tie04} -- 855\,kpc and 867\,kpc, respectively. However, the detached eclipsing binary method gives a significantly longer distance of 964\,kpc \citep{bon06}, while a cepheid based distance is shorter -- 802\,kpc \citep{lee02}. Therefore, further in this paper we will
use the distance modulus of 24.75, which places M\,33-EC1
at a projected distance of 12.5\,kpc from the M\,33 center.

To estimate the cluster's age and metallicity we compared the
shape and slope of the observed RGB with the isochrones of \citet{gir02} and \citet{van06}. In the case of \citet{gir02} isochrones, we achieved the best fit for the interpolated isochrone of the age of 13\,Gyr and metallicity of [M/H]\,=\,-1.2 (Fig.\,2, panel a; isochrone of the age of 14\,Gyr and metallicity of [M/H]\,=\,-1.3
is over-plotted). However, the isochrone of lower metallicity,
[M/H]\,=\,-1.4, and age of $\sim$18\,Gyr, as well as the isochrone
of higher metallicity, [M/H]\,=\,-1.0, and age of $\sim$2.5\,Gyr,
can also be fitted reasonably well. Therefore, additional
information is needed in order to break the age (2.5--18 Gyr) and
metallicity ([M/H]\,=\,-1.4 -- -1.0) degeneracy.

We obtained more constrained fits by employing \citet{van06}
isochrones (2--18\,Gyr), which are available on the finer
metallicity grid for three alpha element abundance ratios
([$\alpha$/Fe]\,=\,0.0, 0.3, 0.6). We achieved good, although
degenerate, fits for the ages of $>$7\,Gyr and metallicity of
[M/H]\,$<$\,-1.4 independent on alpha element abundance, see
Fig.\,4. The isochrones spanning a narrow age and metallicity
range are over-plotted on CMDs for the illustrative purpose.
Assuming the reasonably old cluster's age of 13\,Gyr,
we derived metallicity of [M/H]\,=\,-1.6. Note, that for the same
age (13\,Gyr) metallicity derived from the \citet{gir02}
isochrones is higher by 0.4\,dex. The derived metallicity
[M/H]\,$\lesssim$\,-1.4 is in good agreement with the recent
spectroscopic metallicity determination of the M\,33 halo stars in
a nearby field to the M\,33-EC1 location \citep{mcc06}.

\subsection{Integrated photometry and structural parameters}
The M\,33-EC1 age of $>$7\,Gyr, estimated from the isochrone
fitting, implies that it should possess a globular cluster-like
surface number density profile, which is traditionally fitted
by the King model \citep{kin62}:

\[\rho(r)=\rho_0\cdot[(1+(r/r_{\rm c})^2)^{-1/2}-(1+(r_{\rm t}/r_{\rm
c})^2)^{-1/2}\,]^2,\]

\noindent where $\rho_0$ -- central surface number density, $r_{\rm c}$ -- core radius, and $r_{\rm t}$ -- tidal radius are profile
fitting parameters. However, due to a small number of bright stars and an incompleteness of the stellar photometry catalogue at fainter magnitudes (see Fig.\,3), we decided to fit the King model to the surface brightness, rather than to the surface number density profile. This assumption is reasonable for the surface brightness profiles, constructed from aperture photometry, which even at large radial distances sample the cluster's stellar population satisfactorily well.

On the other hand, large M\,33-EC1 extent and relatively low
luminosity (mass), as well as long ($\sim$12.5\,kpc) projected
distance from the host galaxy's center, can lead to an assumption,
that the cluster is dynamically young and possesses a surface
brightness profile, which could be reproduced by the empirical EFF
\citep{els87} profile derived for young ($<$300\,Myr) Large
Magellanic Cloud clusters. We employed the EFF model in
differential form representing surface brightness profile

\[\mu(r)=\mu_0\cdot(1+(r/r_{\rm e})^2)^{-n},\]
\noindent and in integral form representing integrated luminosity
profile

\[\Sigma(r)=\Sigma_0\cdot r_{\rm e}^2/(n-1)\cdot[1-(1+(r/r_{\rm e})^2)^{1-n}],\]

\noindent where $\mu_0$ -- central surface brightness, $\Sigma_0$
-- central luminosity, $r_{\rm e}$ -- scale-length, and $n$ --
power-law index.

Determination of an accurate center of the well resolved cluster
is a sensitive procedure in constructing the surface brightness
profile. In the central part of M\,33-EC1 luminous stars are
distributed slightly asymmetrically (see Fig.\,1), therefore, the
location of the center was derived by fitting the luminosity
weighted and spatially smoothed surface brightness profile in the
area of 20\arcsec$\times$20\arcsec. M\,33-EC1 exhibits perfectly
round isophotes at large radii, suggesting that the mass
distribution is spherical in general -- hence we used circular
apertures for the construction of surface brightness and
integrated luminosity profiles, by integrating in 0\farcs4 wide
annuli up to the radius of 20\arcsec.

The sky background was determined in a circular annulus centered
on the cluster and spanning the radial range from 20\arcsec\ to
30\arcsec\ (see Fig.\,1). The correct sky background subtraction
is critical for determining a shape of surface brightness profile
in the cluster's outer region (for detailed discussion see
\citealp{hil06}), therefore, systematic sky background subtraction
errors were evaluated by constructing the profiles with
over-subtracted and under-subtracted sky background. The sky
background variation by R.M.S. of the sky background value,
led to insignificant structural parameter changes.

Resultant M\,33-EC1 radial surface brightness profiles are
smooth up to $\sim$16\arcsec\ radius, where a bright RGB star
is located. Therefore, we performed the King and EFF model profile fitting up to the radius of 14\arcsec. The best model fits to the $V$-band sky background-subtracted integrated luminosity (top panel) and surface brightness (bottom panel) radial profiles are presented
in Fig.\,5.

Directly fitted and derived ($r_{\rm h}$ and full-width at half
maximum, FWHM) M\,33-EC1 structural parameters, as well as
corresponding standard deviations, are listed in Table\,1. Derived
parameters were computed basing on transformation equations (6, 7,
9, 10) presented by \citet{lar06} for the EFF profile

\[{\rm FWHM}=2\cdot\,r_{\rm e}\cdot\,\sqrt{2^{1/n} - 1},\]
\[{r_{\rm h}}=r_{\rm e}\cdot\,\sqrt{0.5^{1/(1-n)} - 1},\]

and the King profile

\[{\rm FWHM}=2\cdot\,r_{\rm c}\cdot\,\sqrt{((1-\sqrt{0.5})/\sqrt{1+(r_{\rm t}/r_{\rm c})^2}+\sqrt{0.5})^{-2}-1},\]
\[{r_{\rm h}}=0.547\cdot\,r_{\rm c}\cdot\,(r_{\rm t}/r_{\rm c})^{0.486}.\]

For further discussion we choose the conservative lower limit of the
cluster's half-light radius of $r_{\rm H}=4\farcs7$, that, at the
estimated distance of 890\,kpc, converts to $\sim$20.3\,pc,
revealing the extended M\,33-EC1 nature. It is also important to note,
that the differences between $V$-, $R$-, $I$-band profile fit
parameters are smaller than their standard deviations. Therefore,
in Table\,1 we give averaged parameters for the three passbands.
It is worth noting, that a change of the fitting radius from 12\arcsec
to 19\arcsec does not influence the derived cluster parameters significantly. Integrated magnitudes and colors derived at the cluster's center and radii of (1 -- 4)$\cdot r_{\rm H}$ are listed in Table\,2. We find no significant color gradient over the entire radial cluster's extent. $V-I$ color of M\,33-EC1, taking into account that
only foreground extinction is present at this galactocentric
distance, is in the color range of the intermediate and old age
($\gtrsim$5\,Gyr) M\,33 clusters \citep{sar07}.

\section{Discussion}

We report a discovery of an extended globular-like star cluster
(eGC) at the outer edge of the M\,33 galaxy, M\,33-EC1. All the previously known clusters in M\,33 are compact ones with core radii of $r_{\rm c}\lesssim2$\,pc (\citealp{cha99}, 2001). Therefore, M\,33-EC1
with $r_{\rm c} \sim 25$\,pc is of a very rare type, and the only
such object found in the Subaru Suprime-Cam wide-field survey frames
($\sim$1$\fdg$1$\times$1$\fdg$7) of the M\,33 galaxy.

The $r_{\rm h}$ -- $M_V$ diagram was proven to be a very
informative and suggestive tool for a star cluster study
\citep{vdb04}. In Fig.\,6 we plot this diagram, taking
representative objects from various studies published recently,
and mark M\,33-EC1. The MW galaxy has ten exceptionally large ($r_{\rm
h}\gtrsim15$\,pc) globular clusters \citep{har96}. However, only
two of them (NGC\,5053 and NGC\,2419) are of comparable luminosity or brighter ($M_V$\,$<$\,-6.5) than M\,33-EC1. Recently, four clusters of such extreme-type have also been found in the vicinity of M\,31 \citep{hux05, mac06}. \citet{hux05} pointed out, that eGCs in the MW are fainter than those in the M\,31 galaxy, due to differing formation and evolution scenarios of the host galaxies.

By means of luminosity, structural parameters, and metal-poor
nature, M\,33-EC1 is very similar to NGC\,5053 (MW eGC) and to
four M\,31 eGCs. Therefore, regardless of the difference in
morphological type, size, and luminosity, in the vicinity of
three different galaxies eGCs of the same type reside.
Similar eGCs discovered in the spirals M\,51 and M\,81
\citep{cha04}, and in the giant elliptical galaxy NGC\,5128
\citep{gom06} expand further the variety of eGC's host galaxies
(Fig.\,6). To our knowledge, M\,33 is the smallest spiral galaxy
hosting eGC.

We note that recent studies \citep{sak06, bel07, irw07} do not
warrant a simple classification of stellar systems by using their
structural parameters, therefore, structural parameters of M\,33-EC1
may also be shared with low surface brightness dwarf galaxies. In
this context, the extended nature and very low concentration
($r_{\rm t}/r_{\rm c} \sim 2.5$; Table 1) of M\,33-EC1 suggests that this stellar system could be a low surface brightness dwarf galaxy. The central surface brightness of
$\mu_{0,V} \sim 23$\,mag$\cdot$arcsec$^{-2}$ is both consistent
with lower end of surface brightness of the MW globular clusters
\citep{har96} and with local dwarf galaxies \citep{mat98}. At present,
we have no clear diagnostics to discriminate between these possibilities. The best way to constrain the origin of M\,33-EC1 would be to conduct a study of cluster dynamics. The velocity dispersion data, in particular, would make it possible to determine whether this cluster contains dark matter or not \citep{ben92}, since low surface brightness dSphs are found to exhibit very large mass-to-light ratio, see e.g. \citet{kle05}, \citet{mar07}.

The extended nature of M\,33-EC1 becomes very important when it is
considered in light of merging history of M\,33. Basing on the
assumption that M\,33 has no (prominent) thick disk and/or halo
\citep{fer06}, it has long been postulated that very few, if any,
massive accretion events have taken place. It is still
controversial issue, however, whether M\,33 is a pure stellar disk
system or it has a thick disk and/or a halo component. It is
interesting to note, however, that accumulating evidences suggest
a complex merging history of M\,33. \citet{cha02} reported that
there is a wide spread in the age of star clusters and some old
clusters have velocities consistent with the halo component
dynamics. \citet{mcc06} suggested a halo component and a possible
stream by means of spectroscopic study of individual RGB stars.
The discovered M\,33-EC1 cluster may also give support for the merging
history scenario -- it could be a stripped dwarf galaxy that has
accreted and merged onto M\,33 -- a scenario suggested for eGCs
in M\,31 \citep{hux05}. We note that the proximity of M\,33-EC1
relative to the stream \citep{mcc06} could suggest a physical
connection, however, the preliminary metallicity estimates for both parties differ significantly. We here just mention that the metallicity of globular clusters in Sagittarius dwarf spheroidal do not necessarily agree with that of parent galaxy \citep{bel03}.

\citet{van04} discovered an extended halo in the dwarf irregular
galaxy Leo~A and suggested that even such a small dwarf galaxy was
formed in a much more complex way than believed before, implying
hierarchical galaxy formation on all scales. Recent findings
indicate that small late-type disk galaxies, such as M\,33, could
have experienced merging events. Therefore, the eGC presented in
this paper, M\,33-EC1, together with various objects, which are
suggested to associate with the M\,33 halo, are all important
targets for detailed study in order to understand the merging
history of not only M\,33, but of galaxies on all scales.

\acknowledgments

We are indebted to Chisato Ikuta for her invaluable help with
observations on Subaru telescope. We are thankful to the anonymous
Referee for constructive suggestions and proposed corrections.
This work was financially supported in part by a Grant of
the Lithuanian State Science and Studies Foundation, and by a
Grant-in-Aid for Scientific Research by the Japanese Ministry
of Education, Culture, Sports, Science and Technology (No. 19540245).

\clearpage

\begin{table}
\caption{Structural parameters of the star cluster M\,33-EC1.} \label{tab:1}
\begin{center}
\begin{tabular}{lccccccc}
\hline
 Models  &    $r_{\rm e}$        &    $n$         &   FWHM      &   $r_{\rm h}$\\
\hline
EFF (integral)  &  8.5$\pm$0.8 &  3.6$\pm$0.6 & 7.9$\pm$0.1 & 4.7$\pm$0.2  \\
EFF (differential)  & 10.2$\pm$1.3 &  4.6$\pm$1.0 & 8.3$\pm$0.2 & 4.7$\pm$0.1  \\
\hline
         &  $r_{\rm c}$ &  $r_{\rm t}$ &   FWHM      &   $r_{\rm h}$\\
\hline
King (differential) &  5.8$\pm$0.3 & 14.6$\pm$0.8 & 8.2$\pm$0.2 & 4.9$\pm$0.1  \\
\hline
\end{tabular}
\end{center}
{\it Note}. All parameters are given in arc-seconds.
\end{table}

\clearpage

\begin{table}
\caption{Photometric parameters of the star cluster M\,33-EC1.}
\label{tab:2}
\begin{center}
\begin{tabular}{ccccccc}
\hline
 $r$ & $V$ & $V-R$ & $R-I$ & $M_V$ & $\Sigma_V$ & $\mu_V$ \\

[$r_{\rm H}$] & & & & & [L$_{V, \sun}$$\cdot$pc$^{-2}$] & [mag$\cdot$arcsec$^{-2}$] \\
\hline
0 &  --   & 0.47 & 0.44 &   --  & 22.6 & 23.03 \\
1 & 19.11 & 0.47 & 0.48 & -5.83 & 14.4 & 23.52 \\
2 & 18.50 & 0.47 & 0.47 & -6.44 &  6.3 & 24.42 \\
3 & 18.39 & 0.48 & 0.46 & -6.55 &  3.1 & 25.19 \\
4 & 18.33 & 0.49 & 0.46 & -6.61 &  1.8 & 25.75 \\
\hline
\end{tabular}
\end{center}
{\it Note}. Distance from the cluster's center, $r$, is given in
cluster's half-light radius, $r_{\rm H}=4\farcs7$, units.
\end{table}

\clearpage

\begin{figure}
\epsscale{1.0} \plotone{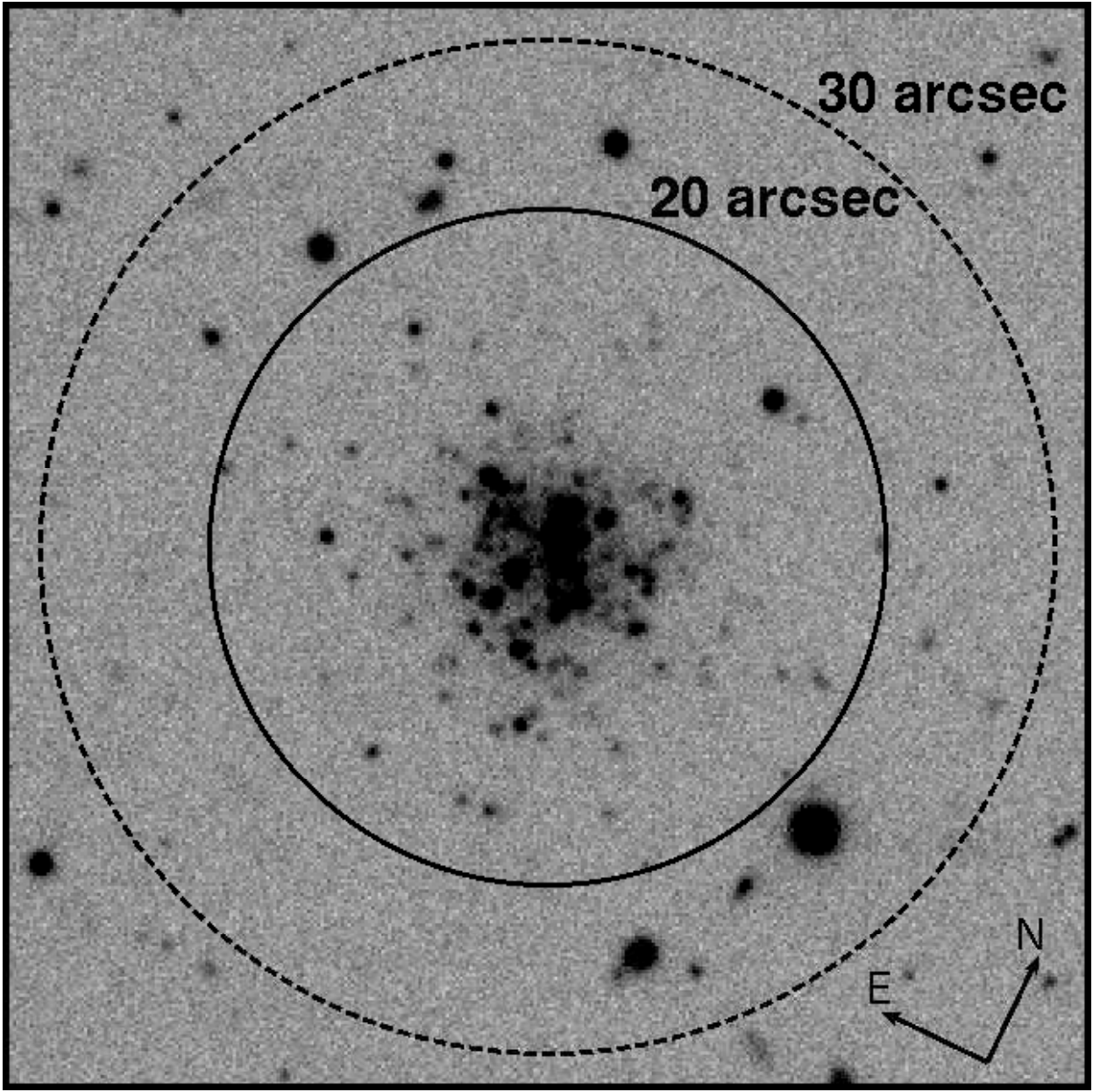} \caption{Suprime-Cam $R$-band
image of the star cluster M\,33-EC1. The circles indicate radii of
20\arcsec\ (solid line) and 30\arcsec\ (dashed line), delineating
the cluster and the sky background determination areas,
respectively. \label{fig1}}
\end{figure}

\clearpage

\begin{figure}
\scriptsize \hspace*{-0.8cm} \epsscale{1.0} \plotone{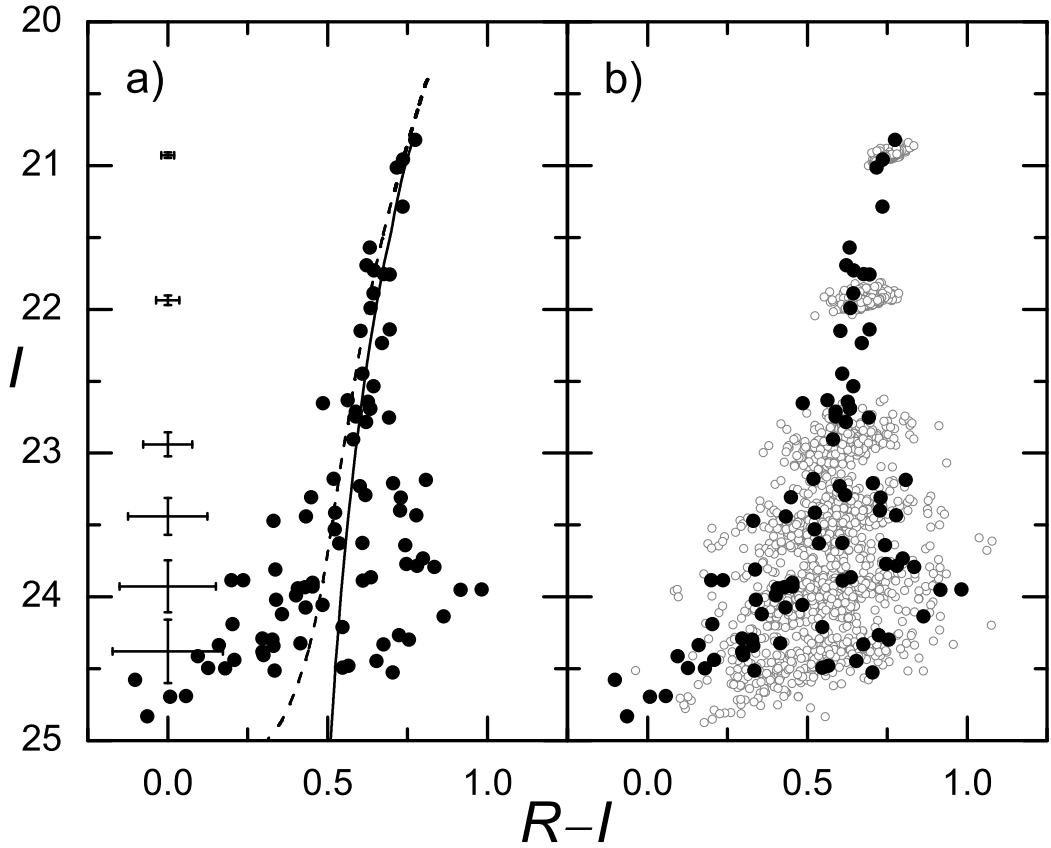}
\caption{The color-magnitude diagram of stellar-like objects
residing in the region of 20\arcsec\ radius from the M\,33-EC1
cluster's center: panel a) overlaid with \citet{gir02} isochrone
of 14\,Gyr and [M/H]\,=\,-1.3, shifted for distance modulus of
24.75, and reddened according to the foreground MW extinction $A_I=0.11$,
$E(R-I)=0.045$, (RGB -- solid line, AGB -- dashed line); panel b)
underlaid with the artificial stars (open circles). Error bars
shown in the panel a) are derived basing on the artificial star
photometry data. \label{fig2}}
\end{figure}

\clearpage

\begin{figure}
\scriptsize \hspace*{-0.8cm} \epsscale{1.0} \plotone{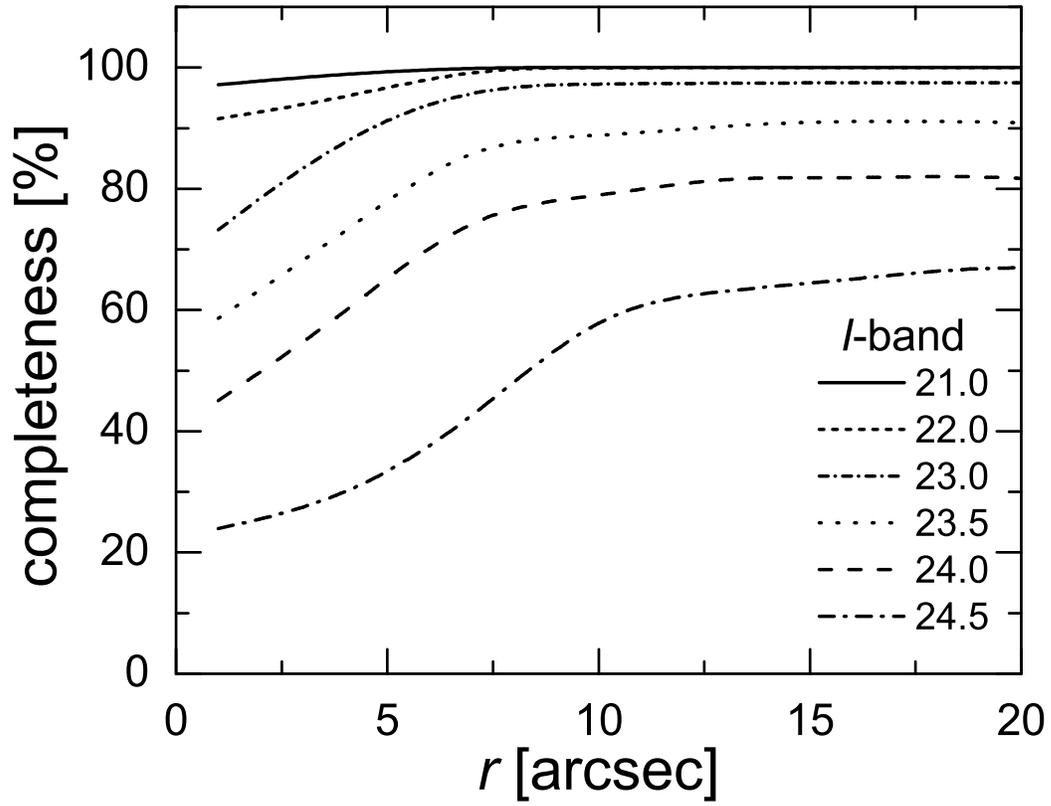}
\caption{Radial dependence of the $I$-band photometry data
completeness at the artificial star test (AST) reference points.
\label{fig3}}
\end{figure}

\clearpage

\begin{figure}
\scriptsize \hspace*{-0.8cm} \epsscale{1.0} \plotone{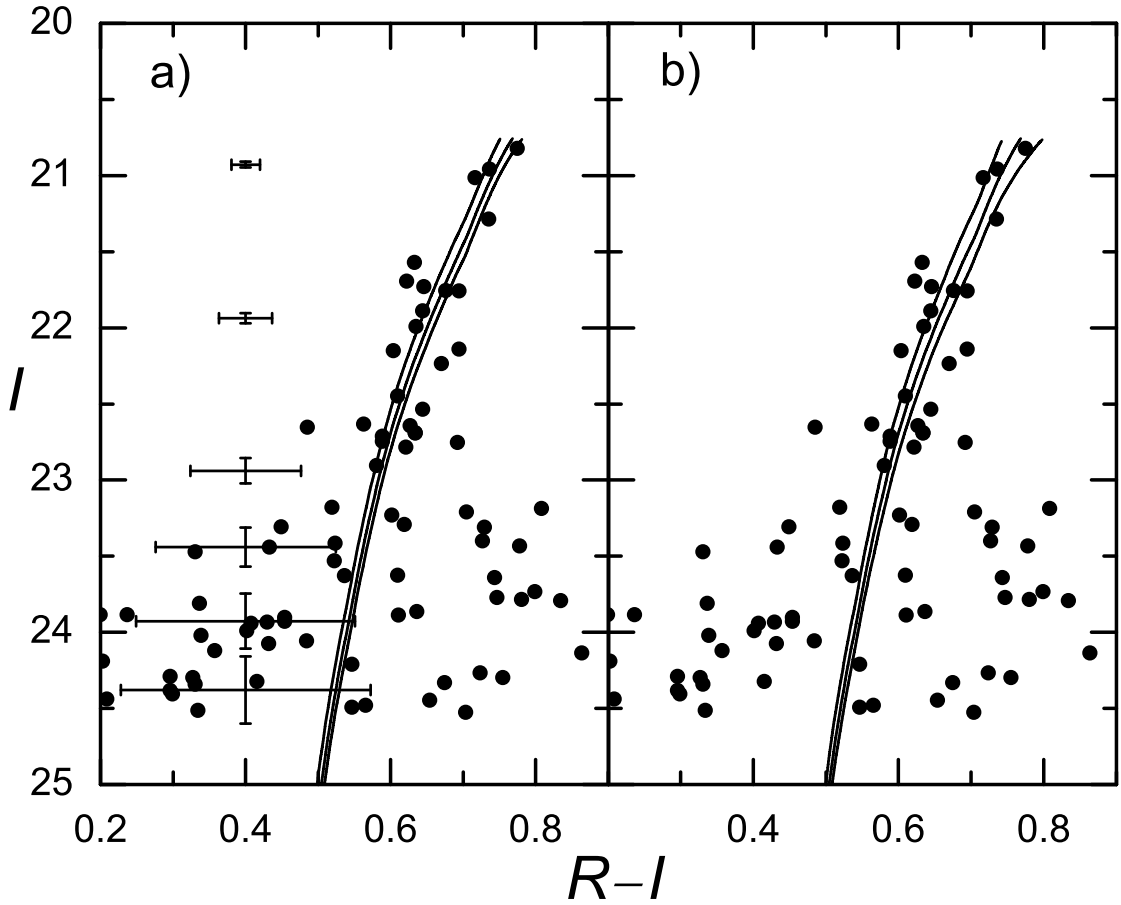}
\caption{The same as in Fig.\,2 panel a) but with \citet{van06}
isochrones overlaid: panel a) metallicity of [M/H]\,=\,-1.53 for
ages 7, 10, \& 13\,Gyr; panel b) age of 10\,Gyr for
[M/H]\,=\,-1.71, -1.53, -1.41. \label{fig4}}
\end{figure}

\clearpage

\begin{figure}
\scriptsize \hspace*{-0.8cm} \epsscale{1.0} \plotone{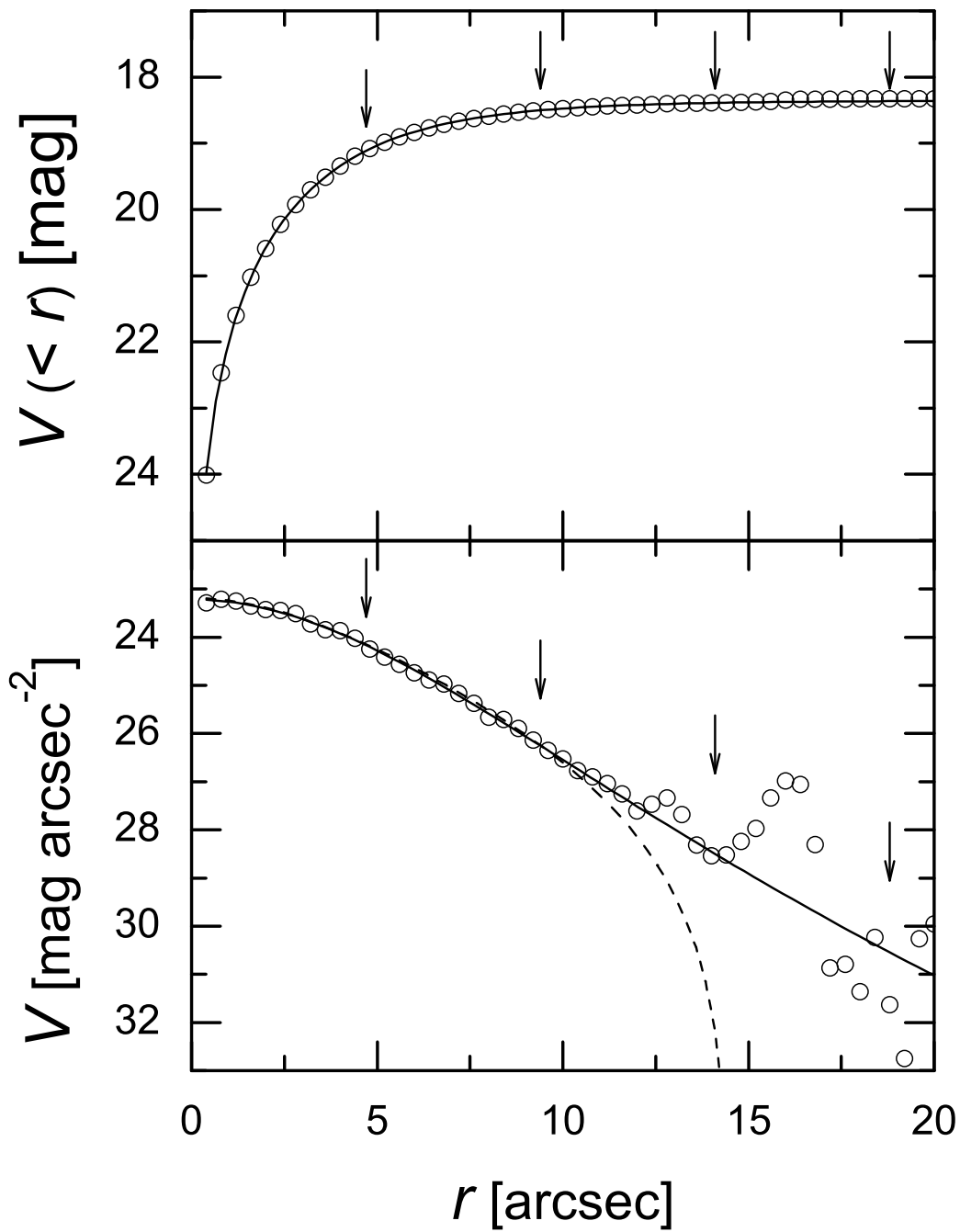}
\caption{The radial profiles of the empirical King (dashed line)
and EFF (solid line) model fits to the $V$-band sky background-subtracted
integrated luminosity (top panel) and surface brightness (bottom panel)
profiles. Small arrows indicate the radii of (1,~2,~3,~4)$\cdot r_{\rm H}$;
$r_{\rm H}=4\farcs7$.
\label{fig5}}
\end{figure}

\clearpage

\begin{figure}
\scriptsize \hspace*{-0.8cm} \epsscale{1.0} \plotone{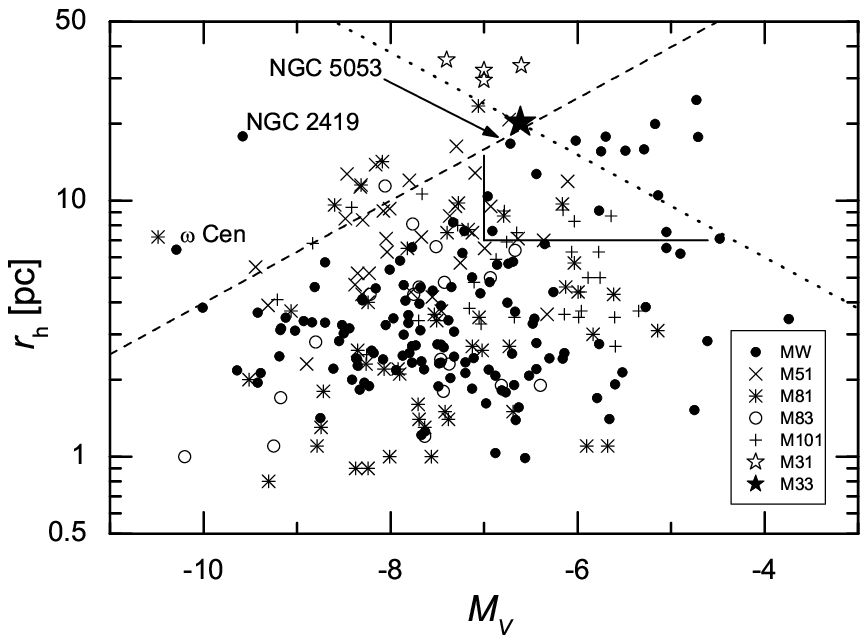}
\caption{Plot of $r_{\rm h}$ vs. $M_V$ for the eGC M\,33-EC1
(filled star). The extended M\,31 clusters \citep{mac06} (open
stars), the MW globular clusters (\citealp{har96}; catalogue
revision: Feb. 2003) (filled circles), and clusters in M\,51
(crosses), M\,81 (asterisks), M\,83 (open circles), M\,101
(pluses) galaxies \citep{cha04} are shown. The star clusters of
M\,33 are not indicated because of their small sizes, $r_{\rm
c}\lesssim2$\,pc (\citealp{cha99}, 2001). Dashed
(log$(r_{\rm h})\,=\,0.2 \cdot M_V+2.6$; \citet{vdb04}) and
dotted (average surface luminosity of
15$\cdot$L$_{V, \sun}$$\cdot$pc$^{-2}$ within $r_{\rm h}$)
lines are drawn for reference. The solid L-shape line marks
a location of faint fuzzy clusters \citep{bro02}. The MW
globular clusters $\omega$\,Cen, NGC\,2419, and NGC\,5053 are
labelled. \label{fig6}}
\end{figure}

\end{document}